\documentclass[fleqn,usenatbib]{mnras}

\usepackage{newtxtext,newtxmath}

\usepackage[T1]{fontenc}

\DeclareRobustCommand{\VAN}[3]{#2}
\let\VANthebibliography\thebibliography
\def\thebibliography{\DeclareRobustCommand{\VAN}[3]{##3}\VANthebibliography}

\usepackage{graphicx}	
\usepackage{amsmath}	
\usepackage{xcolor}

\usepackage{float}
\usepackage{dblfloatfix}

\newcommand{\pcm}{cm$^{-2}$}	
\newcommand{\erg}{erg cm$^{-2}$ s$^{-1}$}	
\newcommand{\src}{4U 1626--67}	
\newcommand{\nus}{\emph{NuSTAR}}	
\newcommand{\astro}{\emph{AstroSat}} 

\usepackage[normalem]{ulem}

\title[QPOs in \src]{Sidebands to mHz QPOs in \src\ in the second spin-down state}

\author[Sharma R. et al.]{
Rahul Sharma,$^{1,2}$\thanks{E-mail: rahul1607kumar@gmail.com}
Chetana Jain,$^{3}$ \thanks{E-mail: chetanajain11@gmail.com}
Biswajit Paul$^{1}$
and Aru Beri$^{4,5}$
\\
\\
$^{1}$Raman Research Institute, C V Raman Avenue, Sadashivanagar, Bangalore 560080, India\\
$^{2}$Inter-University Centre for Astronomy and Astrophysics (IUCAA), Ganeshkhind, Pune 411007, India\\
$^{3}$Hansraj College, University of Delhi, Delhi 110007, India\\
$^{4}$Indian Institute of Science Education and Research (IISER) Mohali, Punjab 140306, India\\
$^{5}$School of Physics \& Astronomy, University of Southampton, Southampton, Hampshire SO17 1BJ, UK}

\date{Accepted XXX. Received YYY; in original form ZZZ}

\pubyear{2025}

\begin{document}
\label{firstpage}
\pagerange{\pageref{firstpage}--\pageref{lastpage}}
\maketitle

\begin{abstract}
We report results from an \astro\ Target-of-Opportunity (ToO) observation of \src, performed on 2023 May 18, soon after the discovery of torque reversal to spin-down in the source. The X-ray emission exhibited significant dependence on both energy and torque state. This work highlights the comparison of timing features of \src\ with a previous \astro\ observation from 2018, when the neutron star was in the spin-up state. The power density spectrum (PDS) of the 2023 observation comprised a sharp peak corresponding to $\nu_{\rm NS}\sim$130 mHz X-ray pulsations along with a prominent quasi-periodic oscillation (QPO) feature at $\nu_{\rm QPO}\sim$46 mHz with $\sim$20\% rms amplitude, which was positively correlated with energy. We also report the detection of sidebands to QPO occurring at a beat frequency ($\nu_{\rm NS}-\nu_{\rm QPO}$) of $\sim$83 mHz with $\sim$8\% rms amplitude, having $>3\sigma$ detection significance. Additionally, we utilized \emph{Nuclear Spectroscopic Telescope ARray} (\nus) observations from the same torque state (2023 May-July) to analogize the presence and energy dependence of sidebands.
The source retains timing properties in this spin-down torque state, similar to those seen in the previous spin-down phase. In sharp contrast, PDS from the 2018 observation was dominated by red noise, an absence of QPOs and a broadening in the wings of the pulse frequency peak, indicating a coupling between periodic and low-frequency aperiodic variability.  Furthermore, we detected the known cyclotron resonance scattering feature (CRSF) at 37 keV in the Large Area X-ray Proportional Counter (LAXPC) spectrum. We explore various mechanisms that could possibly explain the presence of QPOs exclusively during the spin-down state.
\end{abstract}




\section{Introduction}



X-ray pulsars are highly magnetized neutron stars in binary systems, accreting matter from a companion star. Most X-ray pulsars are found in high-mass X-ray binaries (HMXBs) and possess strong magnetic fields of $\gtrsim 10^{11}$ G. As the infalling material is channelled along the neutron star's magnetic field lines, it impacts the magnetic poles, producing periodic X-ray pulsations. A small subset of X-ray pulsars resides in low-mass X-ray binaries (LMXBs), where the companion has a mass of $\lesssim 2 M_{\odot}$. Only five such systems are currently known: Her X--1, \src, GX 1+4, GRO J1744--28, and 4U 1822--37 \citep[see,][]{Bildsten97, Jonker01}. In contrast, the majority of accreting neutron stars in LMXBs have much weaker magnetic fields ($B \lesssim 10^9$ G), insufficient to strongly influence accretion, and are either non-pulsating or observed as accreting millisecond X-ray pulsars (AMXPs). Unlike AMXPs, which exhibit spin periods of a few milliseconds, X-ray pulsars typically have much longer spin periods, ranging from fractions of a second to thousands of seconds.

\src\ is an X-ray pulsar in an ultra-compact LMXB system, which was discovered in 1972 \citep{Giacconi72}, with a short binary orbit of $\sim$42 min and a spin period of $\sim$7.7 s \citep{Rappaport77, Middleditch81, Chakrabarty98, Jain07}. An ultra-compact environment in the presence of a strong magnetic field makes \src\ an interesting source for studying accretion dynamics. The most prominent characteristic of \src\ is its steady spin-up and spin-down episodes on time scales up to several years \citep{Chakrabarty97, Jain09, Camero10, Jain10, Sharma23c}. During 1972--1990, \src\ was steadily spinning up \citep[labelled as 1SU, following the nomenclature of][]{Sharma23c}. A torque reversal to spin-down (labelled as 1SD, hereafter) occurred around 1990 \citep{Chakrabarty97} and lasted around 18 yr. In 2008, \src\ again showed a torque reversal to spin-up \citep[labelled as 2SU, hereafter;][]{Jain09, Camero10, Jain10}. In 2023, the source showed another torque reversal to spin-down \citep[labelled as 2SD;][]{Jenke23, Sharma23c}.

Since its discovery, coherent X-ray pulsations have always been detected in \src. However, all three torque reversals in this source have been accompanied by significant changes in the timing and spectral signatures at X-ray energies. The pulse profile has been strongly dependent on the torque state and energy \citep{Jain10, Beri14, Sharma23c}. This is indicative of changes in the anisotropic radiative transfer within the accretion column in the presence of a strong magnetic field of $\sim10^{12} - 10^{13}$ G \citep{Kii86}. Over the years, the power density spectrum (PDS) has also displayed torque-state dependent occurrence of mHz quasi-periodic oscillations \citep[QPOs;][]{Kaur08, Jain10, Beri14, Sharma23c}. Using \textit{Ginga} data during 1SU, \citet{Shinoda90} reported weak 0.04 Hz QPOs. During 1SD, strong $\sim$48 mHz QPOs were reported from observations made with instruments on-board \textit{Beppo-SAX} \citep{Owens97}, \textit{RXTE} \citep{Kommers98, Chakrabarty98} and \textit{XMM--Newton} \citep{Krauss07}. 
During the spin-up phase, the QPO centroid frequency evolved with time from about 36 mHz in 1983 to about 49 mHz in 1993. But during the subsequent spin-down phase, it decreased at a rate of about 0.2 mHz yr$^{-1}$ \citep{Kaur08}. QPOs were not detected in the PDS during 2SU \citep{Jain10}, however, \textit{XMM--Newton} observation revealed both a 3 and a 48 mHz QPO, though the latter had a low rms amplitude \citep{Beri18}. In the current spin-down episode, $\sim$46 mHz QPOs re-appeared as reported from works based on \emph{Nuclear Spectroscopic Telescope ARray} (\nus) observations \citep{Sharma23c, Tobrej2024}.

The torque reversals in \src\ have also been associated with changes in the X-ray intensity. During 1SU, the source luminosity was about 10$^{37}$ erg s$^{-1}$ \citep{White83}. The X-ray flux decreased during the transition to 1SD \citep{Chakrabarty97}, and the source luminosity increased by a factor of about three during the transition to 2SU \citep{Jain10}. During the current phase (2SD), source luminosity dropped by a factor of $\sim$3 to the luminosities of $\sim4\times 10^{36}$ erg s$^{-1}$ \citep{Sharma23c, Tobrej2024}.

The X-ray continuum of \src\ also correlates with the torque state. The spectrum during 1SU was described with a model comprising of $\sim$0.6 keV blackbody component and power-law component having photon index of $\sim$1 \citep{Pravdo79, Kii86, Angelini95}. The source exhibited a similar spectral shape during the second spin-up phase \citep{Jain10, Camero12, Koliopanos16, Sharma23c}. The energy spectrum was relatively hard in between these states (1SD) \citep{Owens97, Vaughan97, Yi99}.
The spectrum also shows the presence of strong emission lines of highly ionized Ne and O. The iron line is observed to be both torque-state and pulse-phase dependent \citep{Beri15, Beri18, Koliopanos16}.
During the current 2SD, the broad-band spectrum is described by the empirical negative and positive power law with exponential cut-off (\texttt{NPEX}) model along with a soft blackbody temperature of $\sim$0.25 keV and the presence of cyclotron resonance scattering feature (CRSF) at $\sim$36 keV \citep{Sharma23c, Tobrej2024}.

In the current work, we present the results from the timing and spectral analysis of an \astro\ Target-of-opportunity (ToO) observation of \src\ conducted during the 2SD torque state. The results are compared with the previous \astro\ observation of the source made during the 2SU (see Table \ref{tab:obs}). Additionally, we incorporate \nus\ observations for timing analysis.
The paper is organized in the following way. Section \ref{sec:obs} provides a description of the observation and the data reduction procedure. In Section \ref{sec:results}, the results from the timing and spectral analysis of \src\ are presented. Our findings are discussed in Section \ref{sec:diss}.

\section{Observation and Data Reduction}
\label{sec:obs}

\astro\ is India's first multiwavelength astronomical mission, which was launched in September 2015 by the Indian Space Research Organization \citep{Agrawal06}. It comprises of five scientific instruments, four of which can observe a source simultaneously from optical to hard X-ray energy bands \citep{Agrawal06, Singh14}.

The timing and spectral analysis in this work has been performed using data from two observations of \src\ taken with the Large Area X-ray Proportional Counter (LAXPC) onboard \astro\ (see Table \ref{tab:obs}). The first observation (obsID: 9000002100) was conducted on 2018 May 15 (MJD 58253), during the source's spin-up phase (2SU), with a total exposure of 81 ks. The second observation (obsID: 9000005642) was performed on 2023 May 18 (MJD 60082), after the torque reversal to the spin-down phase (2SD), with an exposure of 45 ks.
LAXPC has a timing resolution of 10 $\mu$s, and it consists of three co-aligned proportional counters (LAXPC10, LAXPC20 and LAXPC30) covering the energy range of 3--80 keV and has a total effective area of 6000 cm$^{2}$ at 15 keV \citep{Yadav16, Agrawal17}. In the current analysis, we have used event analysis (EA) mode data from all layers of LAXPC20 only, as LAXPC30 was switched off and LAXPC10 had variable gain due to gas leakage. 
The LAXPC software\footnote{\url{https://www.tifr.res.in/~astrosat\_laxpc/LaxpcSoft.html}} \textsc{LaxpcSoft}: version 3.4.4 was used to process the level 1 data. 
The source and background light curves and spectra were extracted by using the tool \texttt{laxpcl1}. The \texttt{backshiftv3} tool was used to correct the background products for the gain shift. The 2023 data required an additional background down-scaling of 2.4\% to match the source and background data above 50 keV. The light curves were extracted with a binsize of 0.1 s. Solar system barycentre correction was performed by using \texttt{as1bary}\footnote{\url{http://astrosat-ssc.iucaa.in/?q=data\_and\_analysis}} tool by taking JPL DE405 ephemeris and source position RA (J2000) = 16$^{\rm h}$ 32$^{\rm m}$ 16$^{\rm s}$.79 and Dec. (J2000) = $-$67$^{\rm o}$ 27$'$ 39$''$.3 \citep{Lin12}.

\src\ was also observed with the Soft X-ray Telescope (SXT) during both the \astro\ observations in the photon counting mode. But, owing to a low duty cycle ($\sim$20\%) and a time resolution of $\sim$2.3 s, we have not used SXT data in this work.

\begin{table}
    \centering
    \caption{The log of \astro\ observations of \src\ analysed in this work.}
    \resizebox{\columnwidth}{!}{
    \begin{tabular}{cccccc}
    \hline
      Obs-ID    & Date          & Exposure  &   Spin-phase  & Spin Period \\
                & (yy-mm-dd)    & (ks)      &               & (s)    \\
    \hline  
       9000002100 & 2018-05-15 & 81 & 2SU  &   7.670530 (1) \\
       9000005642 & 2023-05-18 & 45 & 2SD   &   7.668064 (5) \\
    \hline     
    \end{tabular}}    
    \label{tab:obs}
\end{table}

\section{Results}
\label{sec:results}
\subsection{Timing Analysis}

We extracted 0.1 s light curves in the 3--30 keV energy range, optimized for maximum signal-to-noise ratio and searched for X-ray pulsations using the epoch folding method \citep{Leahy83}. The spin period determined from both the \astro\ observations is mentioned in Table \ref{tab:obs}. The bootstrap method described in \citet{Boldin13} was used to estimate the error in the spin period by simulating 1000 light curves \citep{Sharma23}. These results are consistent with \emph{Fermi}/Gamma-ray Burst pulsar Monitor\footnote{\url{https://gammaray.nsstc.nasa.gov/gbm/science/pulsars/lightcurves/4u1626.html}} \citep{Malacaria2020}. 

\begin{figure}
\hspace{0.5cm}
	\includegraphics[width=0.8\columnwidth]{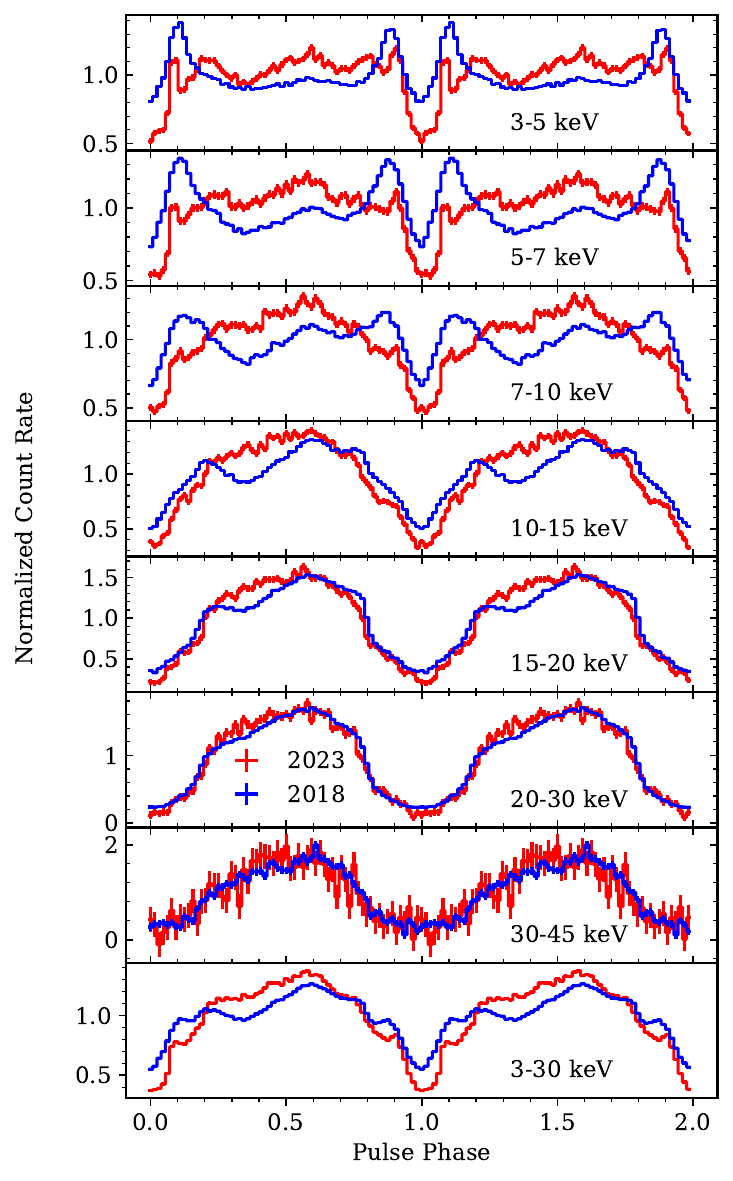} 
    \caption{Energy-resolved pulse profile from \astro\ observations of \src\ taken during 2018 (blue colour) and 2023 (red colour).}
    \label{fig:profile23}
\end{figure}

The 3--30 keV light curves from both \astro\ observations were folded with their corresponding pulse period to obtain pulse profiles (Figure \ref{fig:profile23}). The energy-resolved pulse profiles were extracted in the 3--5, 5--7, 7--10, 10--15, 15--20, 20--30 and 30--45 keV energy ranges, aligned by their minima and overlaid in Fig. \ref{fig:profile23} for comparison. 
At low energies (up to 20 keV), the pulse morphology during 2023 is quite different from that during 2018. At energies up to 5 keV, the 2018 profile shows two peaks separated by a broad, shallow dip covering about 0.5 pulse phase. Then, up to about 10 keV, two shallow dips appear at the pulse phase of 0.4 and 0.7. Thereafter, the profile tends to become sinusoidal. This is consistent with the profiles of the spin-up phase reported by \citet{Beri14} using \textit{RXTE} data and by \citet{Iwakiri19} from \textit{Suzaku} and \nus\ data. The 2023 pulse profile shows several sub-structures up to about 7 keV. Beyond this, the shape is largely sinusoidal, and the width of the main dip increases with energy. This is somewhat similar to profiles seen by \citet{Krauss07} during the first spin-down and by \citet{Sharma23c} during the current phase. We have found that the 3--30 keV pulsed fraction for the 2023 profile is higher ($57.3 \pm 1.3$\%) than that of 2018 profiles ($39.4 \pm 0.4$\%).  

\begin{figure}
	\includegraphics[width=\columnwidth]{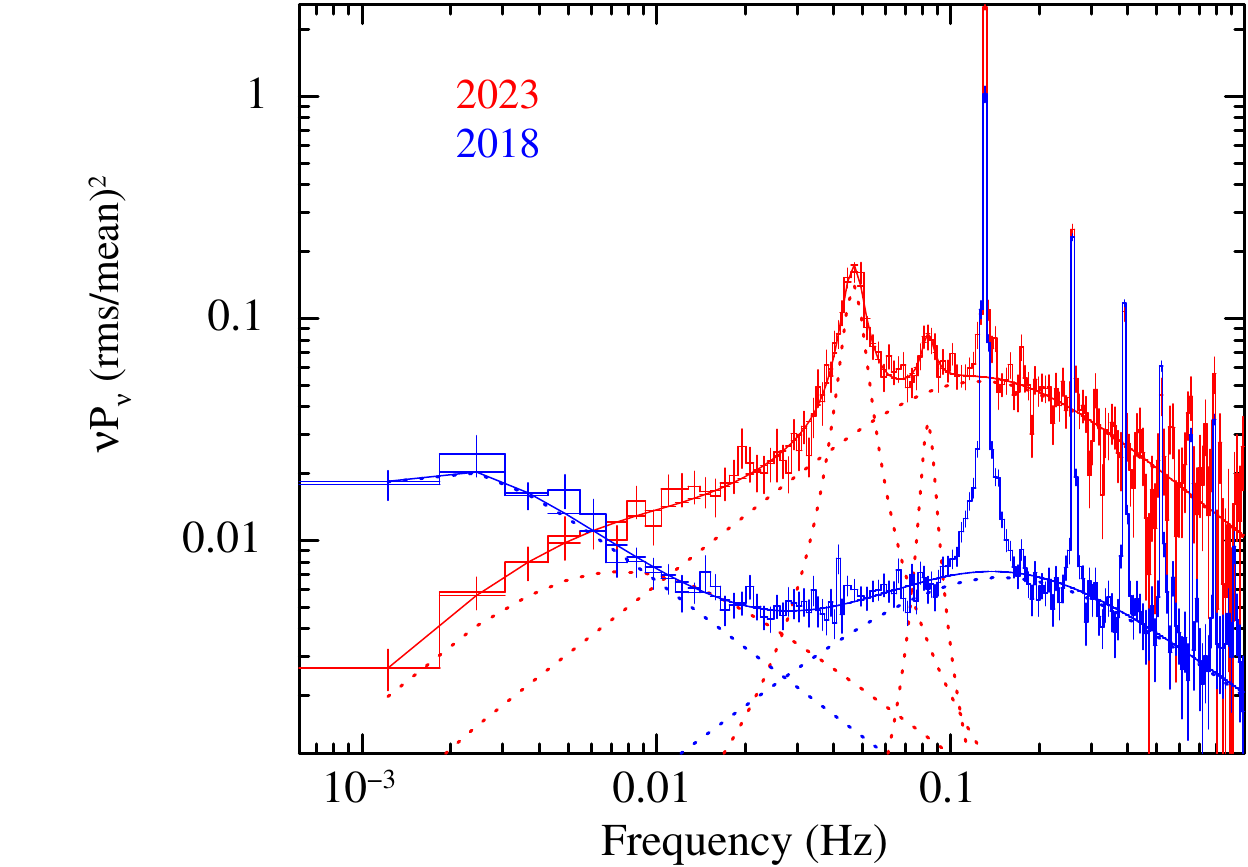}
    \caption{The 3--30 keV PDS of \src\ from \astro\ observations of 2018 and 2023, fitted with a combination of Lorentzian components. Thick lines correspond to the best-fitting model, and dotted lines correspond to individual components of the model. Sharp peaks corresponding to the neutron star's spin period of 7.66 s and its harmonics were omitted during fitting.}
    \label{fig:pds23}
\end{figure}

\begin{table}
    \centering
    \caption{Best-fit of PDS for the 2018 and 2023 \astro\ observations.}
    \resizebox{0.9\columnwidth}{!}{
    \begin{tabular}{cccc}
    \hline
      Component & Parameter    & 2018    & 2023  \\
    \hline
     &    Frequency, $\nu_{0}$ (mHz) & $0^{\rm fixed}$  & $0^{\rm fixed}$ \\
 Lorentzian 1 &      FWHM, $\Delta$ (mHz) & $3.1^{+0.7}_{-0.5}$ & $15.05^{+6.2}_{-4.7}$ \\
     &   rms (\%) & $26.1 \pm 2.1$  & $15.1_{-1.9}^{+2.1}$  \\
\hline
     &   Frequency, $\nu_{0}$ (mHz) & $0^{\rm fixed}$ & $33 \pm 22$ \\
 Lorentzian 2 &   FWHM, $\Delta$ (mHz) & $298^{+18}_{-16}$ & $243_{-20}^{+26}$ \\
       & rms (\%) & $14.6 \pm 0.2$  & $38.2_{-1.8}^{+1.5}$ \\
\hline
       & Frequency, $\nu_{QPO}$ (mHz) & & $46.5 \pm 0.5$ \\
 Lorentzian 3      & FWHM, $\Delta$ (mHz) & & $8.6^{+1.6}_{-1.4}$  \\
        & $Q$-factor & & 5.4 \\
       & rms (\%) & & $20.2 \pm 1.3$ \\
\hline
       & Frequency, $\nu_{QPO}$ (mHz) & & $83.6 \pm 1.8$ \\
 Lorentzian 4      & FWHM, $\Delta$ (mHz) &  & $9.5^{+10.7}_{-4.5}$ \\
         & $Q$-factor & & 8.8 \\
       & rms (\%) &  & $7.8_{-1.6}^{+2.8}$ \\
\hline
      & $\chi^2$/dof & 113.7/108 & 157/123 \\ 
    \hline     
    \end{tabular}}    
    \label{tab:qpo23}
\end{table}

\begin{figure}
	\includegraphics[width=\columnwidth]{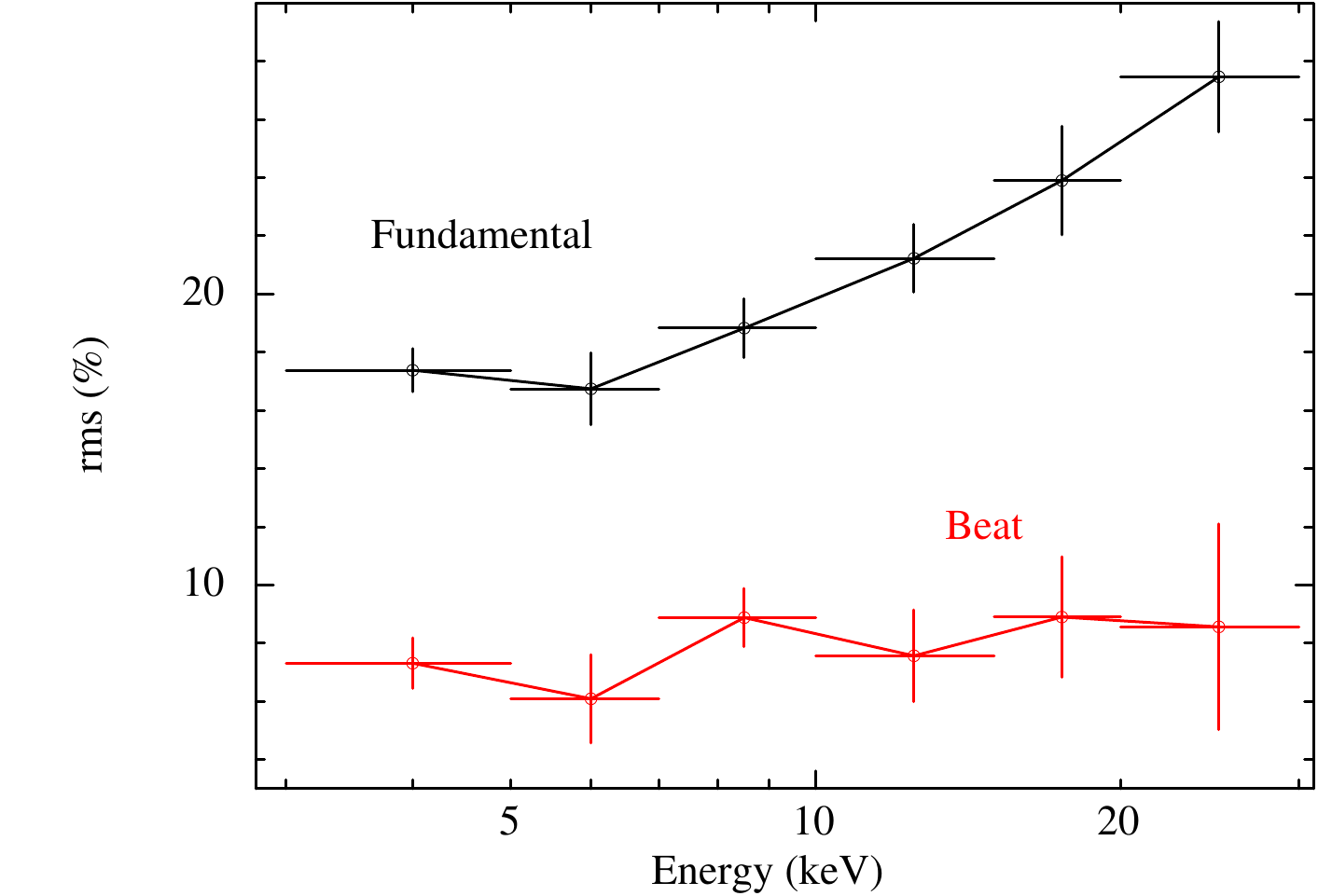 }
    \caption{The variation of rms amplitude of 46 mHz QPO (in black colour) and 83 mHz beat QPO (in red colour) with energy for the 2023 observation (2SD). The rms amplitude of fundamental 46 mHz QPO shows a clear positive correlation with energy. The 83 mHz QPO does not show any energy dependence.}
    \label{fig:rms23}
\end{figure}

Figure~\ref{fig:pds23} shows the 3--30 keV PDS of \src\ generated from 2018 and 2023 data. 
The light curves were divided into 819.2 s segments to calculate the Fourier transform. The resulting PDSs were averaged and rebinned geometrically by a factor of 1.02. The PDS were rms normalized with white noise level subtracted to obtain the rms fractional variability of the time series. Both PDS exhibited sharp peaks at the spin frequency of the pulsar ($\sim$130 mHz) with multiple harmonics and red noise. 
In addition to coherent pulsations, narrow excess in power at $\sim$46 and $\sim$83 mHz were observed in the 2023 PDS, indicating the presence of QPO. The 83 mHz QPO is not a harmonic of the 46 mHz and can be attributed to the beat between the 130 mHz coherent pulsation and the 46 mHz QPO ($\nu_{\rm beat} = \nu_{\rm NS} - \nu_{\rm QPO}$). This is the first-ever detection of 83 mHz beat QPO during the current spin-down phase (2SD) of \src. 

The PDS can be described  with a sum of Lorentzian functions \citep[e.g.,][]{Belloni02, Reig2008, Sharma24}, each defined as
   \begin{eqnarray}
   \label{eqn:lore}
   P(\nu) = \frac{r^2 \Delta}{2 \pi} \frac{1}{(\nu - \nu_0)^2 + (\Delta/2)^2},
   \end{eqnarray}
where $\nu_0$ is the centroid frequency, $\Delta$ is the full-width at half-maximum (FWHM), and $r$ is the integrated fractional rms The quality factor of Lorentzian $Q=\nu_0/\Delta$ is used to validate the presence of a QPO above the noise level. The Lorentzian components with $Q>2$ are generally considered as QPOs, which are otherwise labelled as band-limited noise \citep[e.g.,][]{Belloni02}. The PDS of \src\ from both observations in the 0.001--1 Hz range can be well described with two Lorentzian components representing the red noise continuum. For the 2023 observation, two additional Lorentzians were required to model QPOs at $\sim$46 and $\sim$83 mHz. Modelling of these features yielded a quality factor of 5.4 and 8.8, and rms values of $\sim$20\% and $\sim$8\%, respectively. During the fitting, data points around the fundamental spin frequency and its harmonic were excluded. The 83 mHz QPO was detected with a 3.6 $\sigma$ significance, calculated by dividing the normalization of the Lorentzian component by its 1 $\sigma$ lower bound. We also confirmed a false alarm probability of $<10^{-4}$ using \texttt{simftest}, corresponding to $>$3 $\sigma$ significance.
Table~\ref{tab:qpo23} summarizes the best-fitting parameters of PDS for both observations.
Notably, the red noise component showed significant differences between the two epochs: the 2018 observation exhibited higher power at low frequencies and a broader pulse frequency peak compared to 2023. The low-frequency noise component had an rms of $\sim$26\% in 2018 and $\sim$15\% in 2023, while the high-frequency noise component showed an inverse trend, with $\sim$15\% rms in 2018 increasing to $\sim$38\% in 2023. This suggests a notable evolution in the source's timing properties between the two epochs.

Figure~\ref{fig:rms23} shows the variation of the rms amplitude of 46 mHz QPO and 83 mHz beat QPO with energy for the 2023 observation. The horizontal error bar corresponds to the energy ranges used to generate the PDS, while the vertical error bars show the 1$\sigma$ error in the rms amplitude. The rms amplitude of fundamental 46 mHz QPO shows a clear positive correlation with energy. In contrast, the 83 mHz QPO shows no significant energy dependence.

\begin{figure}
	\includegraphics[width=\columnwidth]{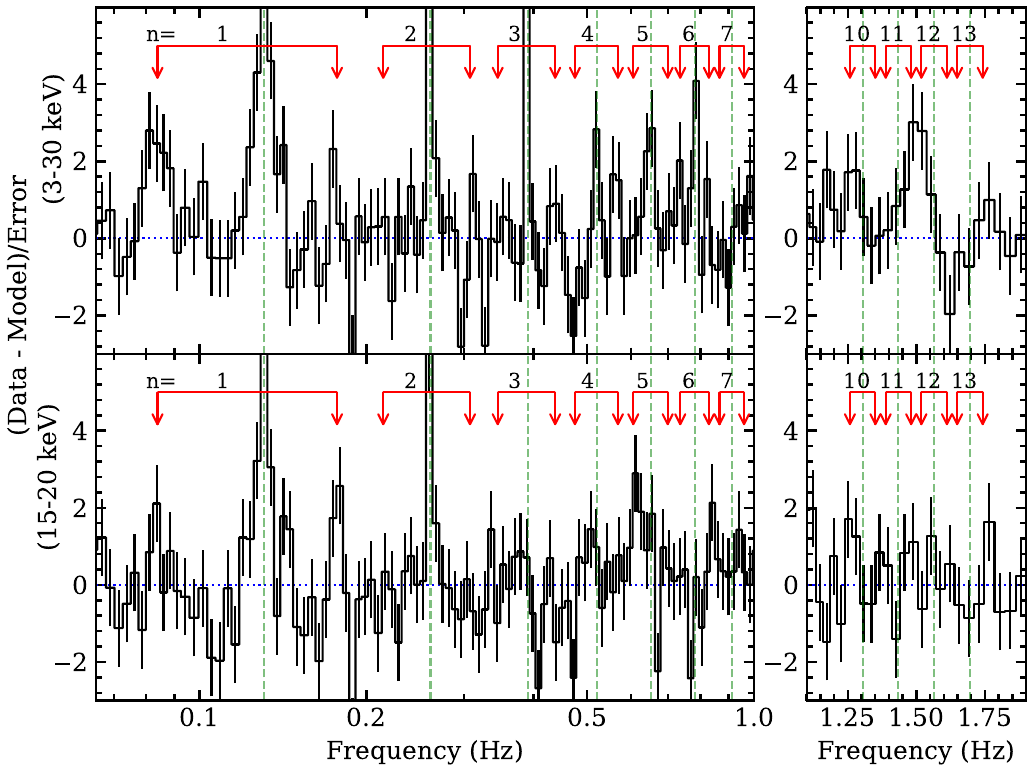 }
    \caption{Residuals from the PDS in the 3-30 keV (top) and 15--20 keV (bottom) bands after modelling of the red noise and the 46 mHz QPO. The sharp peaks correspond to the coherent pulsations and their harmonics, marked by green dashed lines with orders ($n\nu_{\rm NS}$, where $n$=1, 2, 3,...). The pair of red arrows mark the position of symmetric sidebands (lower and upper sidebands at $\nu_{\rm ls}=n \nu_{\rm NS} - \nu_{\rm QPO}$ mHz and $\nu_{\rm us}=n \nu_{\rm NS} + \nu_{\rm QPO}$ mHz, respectively) as if the QPO signal modulates the amplitude of the coherent pulsations. The right panel shows the residuals in the rescaled frequency range, highlighting a strong 1.5 Hz feature between the $n$=11 and 12 harmonics in the 3--30 keV PDS.}     
    \label{fig:sideband}
\end{figure}

Fig. \ref{fig:sideband} shows the residuals after modelling the red noise and 46 mHz QPO in the PDS for the 3--30 keV (top) and 15--20 keV (bottom) bands. The fundamental spin frequency and its harmonics are also indicated with their orders (green dashed lines) and the expected sidebands (red arrows) as if the QPO signal modulates the amplitude of the coherent pulsations. In addition to the 83 mHz feature, an upper sideband at 177$\pm$2 mHz ($\nu_{\rm NS} + \nu_{\rm QPO}$) was also apparent in both energy ranges. Energy-resolved PDS analysis revealed that the 177 mHz feature was significant only in the 15--20 keV range, with a detection significance of more than $3 \sigma$ and rms of 10$\pm$3\%. The 15--20 keV PDS also showed additional features at other sideband frequencies, though at lower significance ($<3 \sigma$). For instance, a feature at 615(12) mHz (lower sideband of $n$=5) was detected with $2.5 \sigma$ significance and rms of 12$\pm$4\%. 
Beyond 1 Hz ($n>7$), most sidebands appeared at a low confidence level in both energy bands, with no statistically robust detections, except for a prominent feature at 1.504(4) Hz. This feature was observed only in the 3--30 keV PDS, with a $4\sigma$ significance and rms of 9$\pm$2\%, and was not apparent in any of the energy-resolved PDS. The right panel of Fig. \ref{fig:sideband} shows the residuals near this frequency. A similar QPO at 1.3 Hz was previously observed in HMXB pulsar XTE J0111.2--7317 \citep{Kaur07}, although the underlying origins are likely different. The 1.5 Hz feature in \src\ could result from an overlap between $n$=11 upper sideband and $n$=12 lower sideband or may arise from instabilities in the accretion flow \citep{Li24}.


\subsubsection{NuSTAR PDS}

We also conducted a search for sidebands in the \nus\ \citep{Harrison13} light curves to investigate their origin. The analysis focused on the five \nus\ observations carried out during the 2SD phase (Table \ref{tab:nus}). Details of these observations and the corresponding data reduction processes are outlined in Appendix \ref{appendix:nustar}. Initial results from the spectral and timing analyses of the 2SD phase were reported by \citet{Sharma23c}. Subsequently, \citet{Tobrej2024} presented spectral and timing results from four of the \nus\ observations taken during 2SD. However, these observations have not been previously utilized to identify sidebands in the PDS.

\begin{figure*}
    \includegraphics[width=\columnwidth]{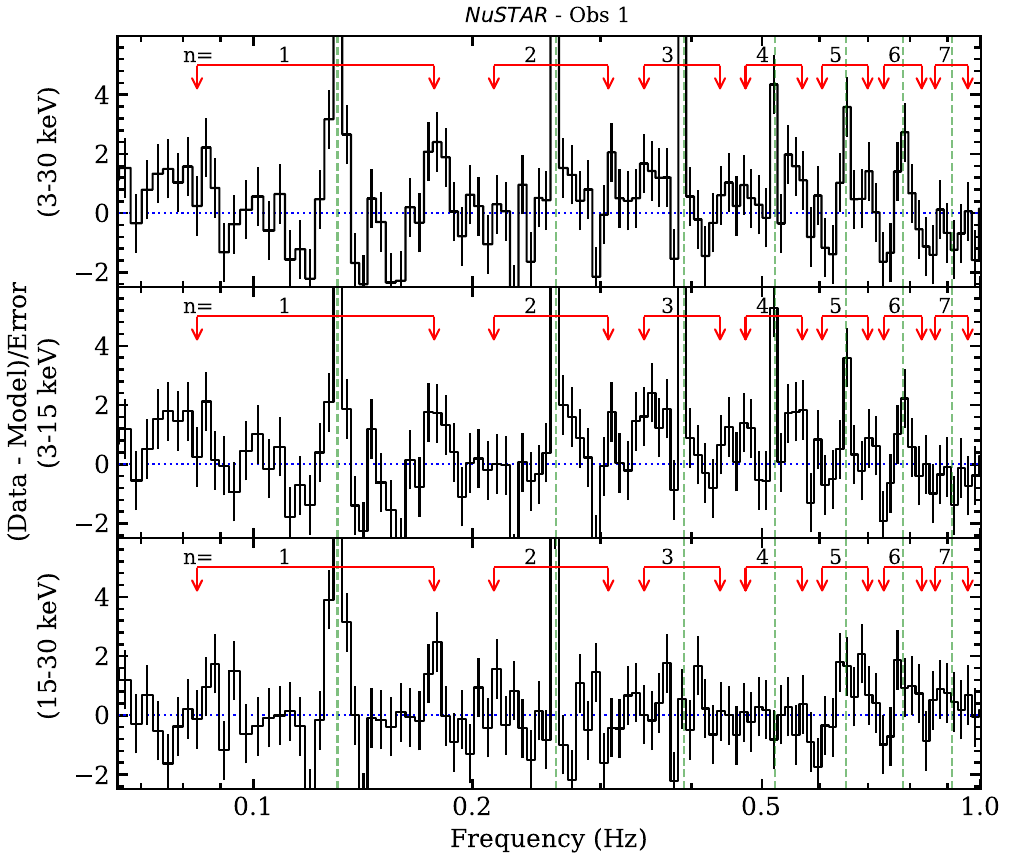}
    \includegraphics[width=\columnwidth]{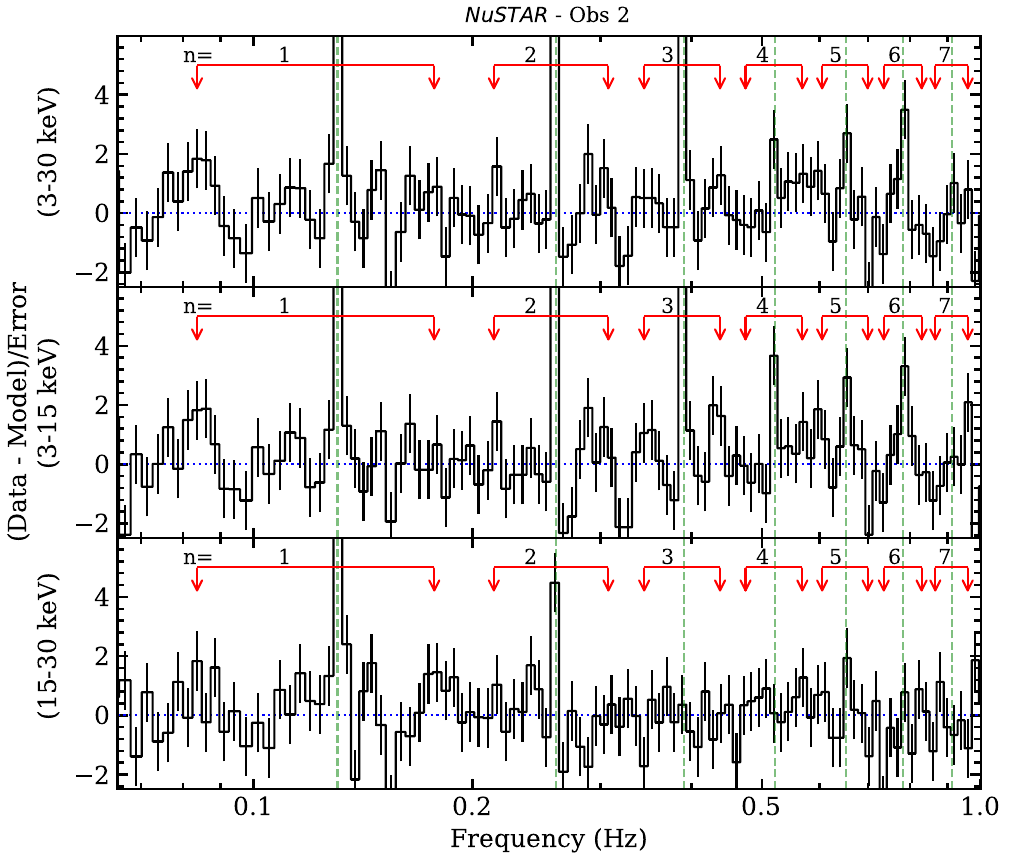}
    \includegraphics[width=\columnwidth]{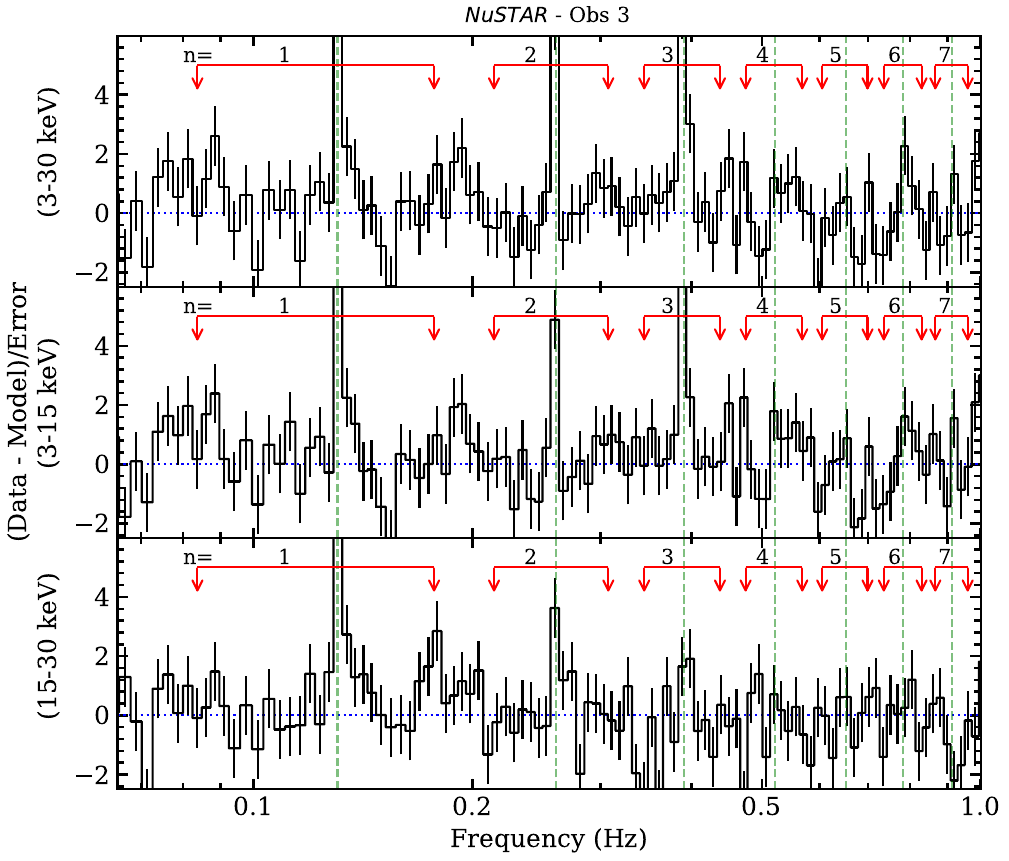}
    \includegraphics[width=\columnwidth]{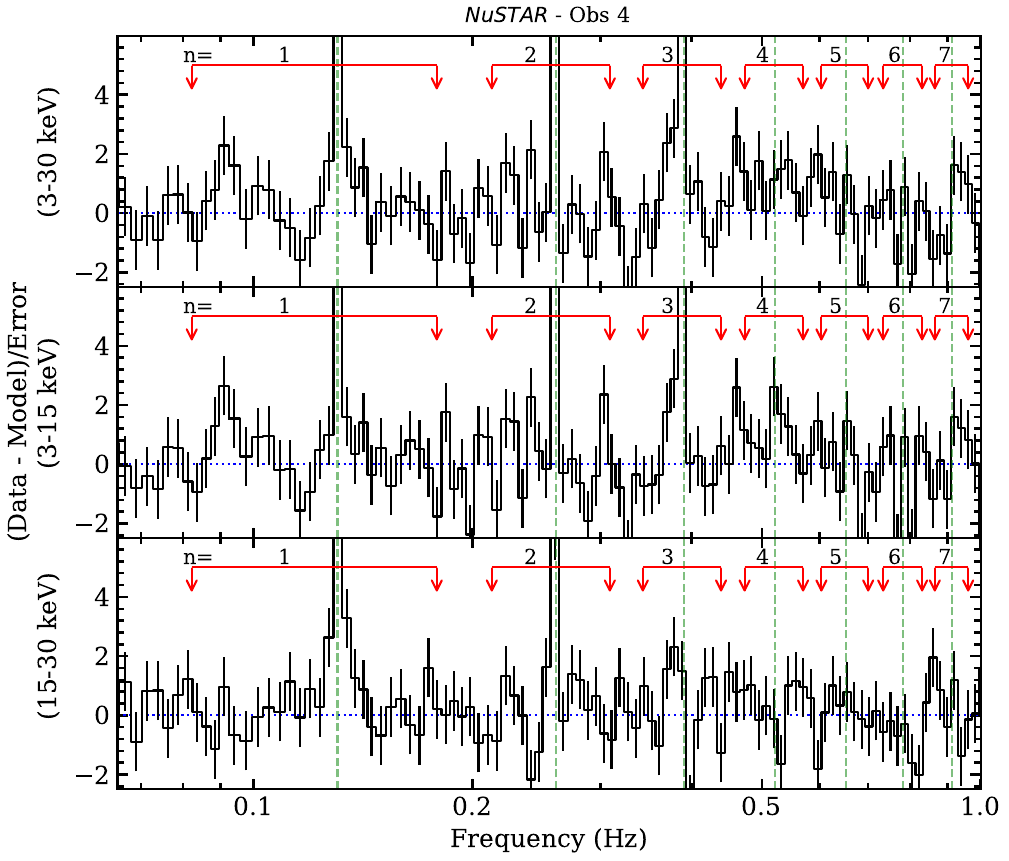}
	\includegraphics[width=\columnwidth]{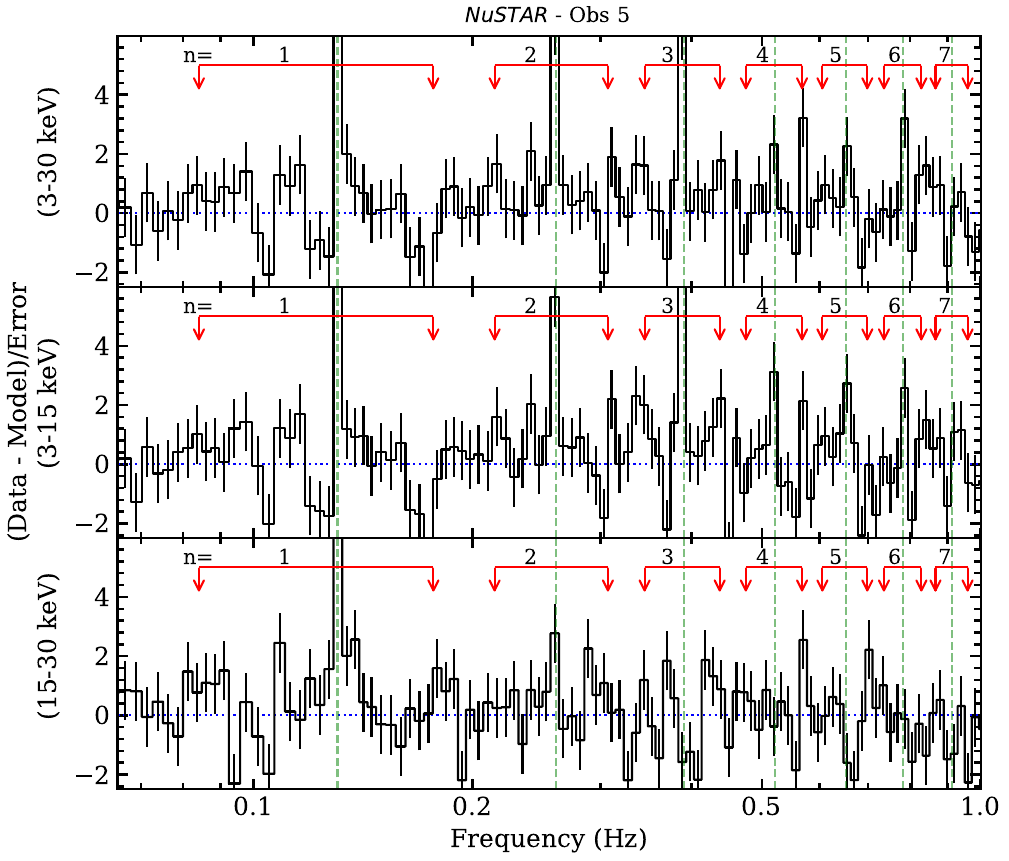}
    \caption{Residuals from the \nus\ PDS in the 3--30 keV (top), 3--15 keV (middle) and 15--30 keV (bottom) bands after modelling of the red noise and the 46 mHz QPO. The sharp peaks correspond to the coherent pulsations and harmonics marked by green dashed lines with orders ($n\nu_{\rm NS}$, where $n$=1, 2, 3,...). The pair of red arrows mark the position of symmetric sidebands: lower and upper sidebands at $\nu_{\rm ls}=n \nu_{\rm NS} - \nu_{\rm QPO}$ mHz and $\nu_{\rm us}=n \nu_{\rm NS} + \nu_{\rm QPO}$ mHz, respectively.}
    \label{fig:nuspds}
\end{figure*}

For comparison with \astro, we extracted PDS in the 3--30 keV (broadband), 3--15 (low energy) and 15-30 keV (high energy) ranges from \nus\ observations. Fig. \ref{fig:nuspds} presents the residuals after modelling the \nus\ PDS by the red noise and 46 mHz QPO. In obs 1, a feature at 177 mHz was clearly detected with $3 \sigma$ significance and rms of 4.6$\pm$1.3\% in the 3--30 keV energy range. This feature was also present in the 3--15 and 15--30 keV range with lower significance, exhibiting rms of $4.2^{+1.0}_{-1.5}$\% and $8.6^{+2.4}_{-3.5}$\%, respectively. Additionally, a lower sideband peaked around 77 mHz and was dominant in 3--15 keV with rms of 7$\pm$2\%.  In obs 2, which was taken just one day before the \astro\ observation, the 83$\pm$2 mHz sideband was detected with rms of 7$\pm$3\%, consistent with \astro\ but at a lower significance of $2.5\sigma$, and only in the 3--15 keV band. No other sidebands could be detected significantly.

In obs 3, no sidebands were significantly detected in the 3--30 and 3--15 keV ranges. However, a feature at 191 mHz was present, albeit with low significance. In the 15--30 keV PDS, the 177(3) mHz feature was detected with $>3 \sigma$ significance and rms of 12$\pm$3\%, consistent with \astro\ PDS in the 15--20 keV range. Additionally, an excess noise component was observed at 19$\pm$2 mHz with rms of $9^{+4}_{-3}$\%, which does not correspond to a sideband of any harmonics. In obs 4 and 5, no sidebands were detected. However, a feature at 92 mHz was identified with rms of $6.1^{+1.3}_{-1.6}$\% and a significance of $3 \sigma$ in obs 4. This frequency is slightly offset from the expected 83 mHz sideband, suggesting the presence of additional oscillations in the source intensity.

\begin{figure}
	\includegraphics[width=\columnwidth]{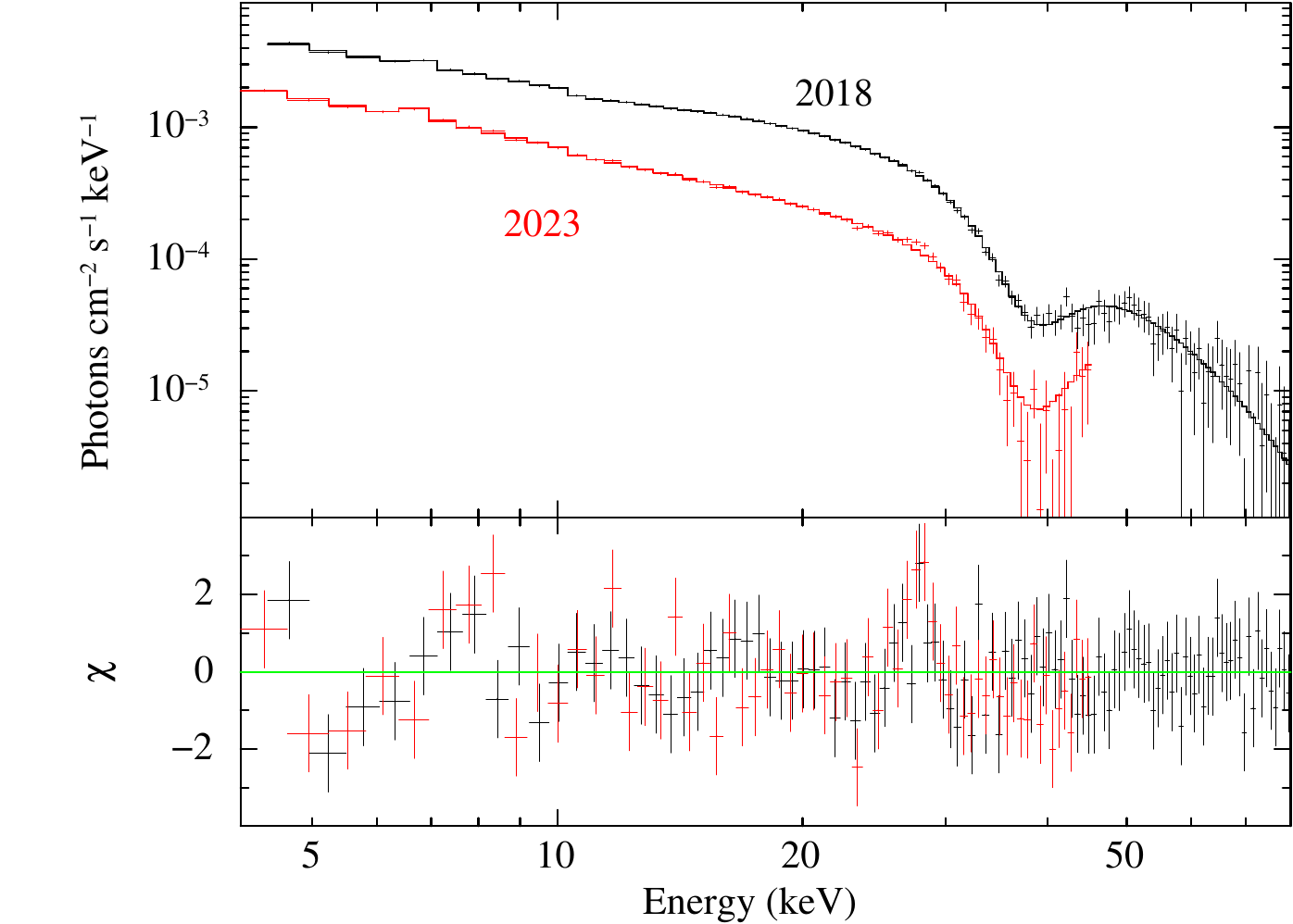 }
    \caption{The best-fitting LAXPC spectrum of 2018 and 2023 observations are shown in the top panel. The bottom panel shows the respective residuals from the best-fitting model. The additional residual near 28 keV can be due to larger uncertainty in background estimation around the K-fluorescence energy of Xe \citep{Antia17, Sharma20}.}
    \label{fig:spec}
\end{figure}

\subsection{Spectral analysis}

To quantify the spectral variation, we performed a spectral analysis using LAXPC data from the 2018 and 2023 observations. A systematic uncertainty of 1.5\% was applied during spectral fitting. For the 2023 observation, we used data up to 45 keV, as the spectrum above this energy was dominated by background noise, while the 2018 spectrum extended up to 80 keV. Both spectra were rebinned using \textsc{grppha} to have a minimum count of 20 counts per bin and fitted both spectra simultaneously with an absorbed \texttt{NPEX} model \citep{Mihara95, Makishima99}, following the methodology of \citet{Sharma23c}. The line-of-sight absorption, modelled with \texttt{tbabs}, was fixed at $N_H = 9.6 \times 10^{20}$ \pcm\ \citep{HI4PI}. 

The absorbed \texttt{NPEX} model failed to provide a satisfactory fit, showing clear residuals around 37 keV, due to CRSF \citep{Orlandini98, Coburn02, Iwakiri19, Sharma23c}. To account for this, we included the \texttt{cyclabs} model, which features a pseudo-Lorentzian optical-depth profile to describe the CRSF \citep{Makishima90, Mihara90}. For the 2023 spectra, the CRSF line energy and width can not be reliably constrained and, hence, were fixed at the values obtained from the 2018 observation. 
Additionally, an Fe K$\alpha$ emission line at 6.75 keV with a fixed width of 0.15 keV was included, following \citet{Sharma23c}, as allowing the line energy to vary resulted in the parameter pegging at 7 keV. An extra edge component at 10 keV \citep{Manikantan23} was also required to fully describe the continuum, yielding a satisfactory fit. 

The best-fitting model with their respective residuals is shown in Fig. \ref{fig:spec}, and the best-fitting spectral parameters are presented in Table \ref{tab:spec}. No significant spectral variability was detected between the two spectra from the different torque states. Although the average unabsorbed flux in the 3--30 keV range dropped from $7.7 \times 10^{-10}$ \erg\ in 2018 to $2.4 \times 10^{-10}$ \erg\ in 2023. This factor of $\sim$3 reduction in flux is consistent with previous findings by \citet{Sharma23c}.

\renewcommand{\arraystretch}{1.1} 
\begin{table}
	\centering
	\caption{Best-fitting spectral parameters of \src\ for the 2018 and 2023 LAXPC observation. All errors and upper limits reported in this table are at a $90\%$ confidence level ($\Delta \chi^2=2.7$).}
	\label{tab:spec}
	\resizebox{0.9\columnwidth}{!}{
	\begin{tabular}{lccr} 
		\hline
		Component & Parameters & 2018 & 2023 \\
		\hline
		tbabs & $N_{\rm H}$ ($10^{20}$ \pcm) & \multicolumn{2}{c}{$9.6^{\rm fixed}$} \\
	             
		NPEX & $\Gamma$ & $0.59 \pm 0.08$ & $0.65 \pm 0.04$ \\
		        & $f$ ($10^{-3}$) & $1.82 \pm 0.28$ & $0.22^{+0.10}_{-0.08}$\\
		        & $E_{\rm cut}$ (keV) & $7.1 \pm 0.4$ & $11.1^{+1.4}_{-1.1}$\\
		        & Norm ($10^{-2}$) & $1.86^{+0.29}_{-0.24}$ & $0.72^{+0.06}_{-0.05}$ \\

        Gaussian &  $E_{\rm Fe~K}$ (keV) & \multicolumn{2}{c}{$6.75^{\rm fixed}$}\\
            & $\sigma_{\rm Fe~K}$ (keV) & \multicolumn{2}{c}{$0.15^{\rm fixed}$}\\
            & Eqw (eV) & $69 \pm 45$ & $123 \pm 52$ \\
            & Norm ($10^{-4}$) & $2.0 \pm 1.3$ & $1.5 \pm 0.5$ \\
          
        Cyclabs & $E_{\rm cyc}$ (keV) & $37.4 \pm 0.4$ & $37.4^{\rm fixed}$ \\
                  & $\sigma_{\rm cyc}$ (keV) & $6.1 \pm 1.0$ & $6.1^{\rm fixed}$ \\
                  & Depth & $1.94 \pm 0.17$ & $2.5^{+0.6}_{-0.5}$ \\

        Edge &  $E_{\rm egde}$ (keV) & $10.4 \pm 0.4$ & $10.3^{+1.0}_{-0.7}$ \\
            & $\tau$ & $0.13 \pm 0.03$ & $0.10 \pm 0.04$ \\

        Flux$^a$   & $F_{3-30~{\rm keV}}$ & $7.7 \times 10^{-10}$ & $2.4 \times 10^{-10}$ \\

		\hline        
		        & $\chi^2$/dof & 80.5/100 & 87.4/56 \\      
		\hline
        \multicolumn{4}{l}{$^a$ Unbsorbed flux in units of \erg.}\\
	\end{tabular}}
\end{table}


\section{Discussion and Conclusions}
\label{sec:diss}

In this work, we have presented the results from the \astro\ observation of the ultra-compact X-ray pulsar \src, made after the recent torque reversal phase (2SU to 2SD) along with a comparison to the previous \astro\ observation from the 2SU phase. We also report results from the timing analysis of \nus\ observations from the 2SD phase. The pulse profile follows the previously known torque-state dependence patterns \citep{Beri14, Sharma23c}. The flux in the 2SD dropped by a factor of three compared to 2SU. The CRSF was detected at 37 keV, consistent with previous reports \citep{Orlandini98, Coburn02, Iwakiri19, Sharma23c}. \src\ is one of the few sources that exhibit both a CRSF and QPOs \citep{Raman21}. This detection adds to the growing list of CRSF observations made with LAXPC \citep[e.g.,][]{Varun19a, Varun19b, Raman21, Devaraj24}. 

The PDS also exhibited torque-state dependence, with notable differences in noise variability and the appearance of QPOs, which were exclusive to the spin-down state \citep[see,][]{Jain10, Sharma23c}. The base of the pulse peak in the PDS from 2SU (2018) shows a broadened shape compared to 2SD (2023), consistent with \citet{Jain10}. This behaviour is similar to X-ray pulsar 4U 1901+03, where the pulse peak narrows when the QPO feature begins to emerge \citep{James2011}. However, it contrasts with IGR 19294+1816, where the QPO feature appears alongside the broadened peak \citep{Raman21}. This broadening of pulse peak arises from the coupling between the coherent periodic pulse variability and the aperiodic red noise component \citep{Lazzati97}. The increased power at low frequencies in the PDS during the spin-up state can be coupled with the pulse variability, leading to the observed broadening. Similar behaviour is observed in 4U 1901+03, where increased low-frequency variability coincides with pulse peak broadening \citep{James2011}. When the QPO appears, low-frequency noise is suppressed, resulting in less broadening of the pulse peak.

A pronounced $\sim$46.5 mHz QPO feature with $\sim$20\% rms and quality factor of $\sim$5 was detected during the 2023 observation (2SD). Additionally, a secondary QPO was observed at $\sim$83.6 mHz, having a quality factor of 8.8 and rms of 7.8\%. 
The appearance of QPOs in X-ray pulsars is often explained by two primary models: the Keplerian Frequency Model \citep[KFM;][]{Klis87} and the Beat Frequency Model \citep[BFM;][]{Alpar85, Lamb85}. As per the KFM, inhomogeneous structures in the Keplerian disc regularly attenuate the pulsar beam. In the case of \src, the option of KFM is ruled out because that requires the neutron star to spin faster than the inner accretion disc ($\nu_{\rm NS} > \nu_{\rm QPO}$), which will result in centrifugal inhibition of mass accretion \citep{Stella86}. 
In the BFM, the QPO is attributed to the modulation in mass accretion rate onto the neutron star's poles at the beat frequency between the spin frequency of the neutron star ($\nu_{\rm NS}$) and the Keplerian orbital frequency ($\nu_{\rm k}$) of the inner accretion disc. According to BFM, $\nu_{\rm QPO} = \nu_{\rm k} - \nu_{\rm NS}$. 
The 46 mHz QPO in \src\ exhibits a torque-state dependence, apparent during the spin-down phase but absent during the spin-up phase. A weak 0.04 Hz QPO was reported during 1SU \citep{Shinoda90}, but it was never detected during 2SU. Therefore, its reappearance during the recent torque reversal to the spin-down state clearly establishes its linkage to the torque state. Since $\nu_{\rm QPO} < \nu_{\rm NS}$, the BFM can provide a plausible explanation for the 46 mHz QPO, corresponding to Keplerian frequency of $\sim$177 mHz and inner disc radius of $5.3\times 10^8$ cm \citep{Manikantan24}.

However, the detection of an additional QPO at 83 mHz during the current 2SD phase with \astro\ adds a layer of complexity to the BFM. This frequency is not a harmonic of the primary 46 mHz QPO. Interestingly, the 83 mHz QPO follows the relationship $\nu_{\rm 83mHz} + \nu_{\rm 46mHz} = \nu_{\rm NS}$, indicative of a beat between the (46 mHz or 83 mHz) QPO and spin of neutron star, or the lower sideband ($\nu_{\rm ls} = \nu_{\rm NS} - \nu_{\rm QPO}$) of the spin frequency. Similar sidebands were observed in the 1SD phase \citep{Kommers98}, caused by the modulation of coherent pulsation by the amplitude of the QPOs. \citet{Kommers98} suggested that these QPOs are produced by a structure orbiting the neutron star at the QPO frequency rather than originating at the inner edge of the accretion disc. However, it is not clear how such a structure could persist in the accretion disc against the differential rotation for more than a decade (during 1SD), disappear during the 2SU phase, and then reappear in the current 2SD phase. If we consider the QPO frequency to represent the inner disc Keplerian frequency $\nu_{\rm QPO} = \nu_{\rm k}$, it implies that the accretion disc lies outside the co-rotation radius ($R_{\rm co}$) of the neutron star. In this scenario, the QPOs could be generated by the interaction between the magnetic field of the neutron star and the inner edge of the accretion disc, where a portion of the disc infiltrates the corotation radius due to a misalignment of the magnetic axis. Notably, the X-ray flux of \src\ has been observed to decrease by a factor of $\sim$3 during the current spin-down phase \citep[see,][]{Sharma23c}, further strengthening the hypothesis of the disc's location outside or close to the corotation radius. 

Other models, like thermal disc instabilities and magnetic disc precession models, are unlikely to account for the observed frequencies, as they typically predict variability of the order of $\sim$1 mHz \citep{Shirakawa02, Roy19, Liu22}. Another promising explanation is the model proposed by \citet{DAngelo10}, which has been used to describe the 40 mHz QPO in Cen X-3 \citep{Liu22}. In this scenario, the disc is truncated just outside $R_{\rm co}$, and centrifugal forces inhibit accretion by transferring angular momentum from the neutron star to the disc, causing the neutron star to spin down. Instead of the mass being ejected, the accreting mass can stay piled up at high surface density in the inner disc, just outside corotation. Over time, mass accumulates in the disc, increasing gas pressure, which eventually pushes the inner edge of the disc inside $R_{\rm co}$, allowing accretion to resume. Once accretion occurs, the gas pressure decreases, and the disc moves back outside $R_{\rm co}$, restarting the cycle. The period of this cycle can range from 0.02 to 20 viscous timescales ($t_{\rm visc}$) of a few hundred seconds \citep{DAngelo10}. This mechanism could provide a reasonable explanation for the observed mHz QPOs in \src\ exclusively to the spin-down torque state.
 
The 46 mHz QPO exhibited strong energy dependence, with its rms increasing with energy \citep[see,][]{Manikantan24}, while the 83 mHz QPO did not show any significant energy dependence (Fig. \ref{fig:rms23}).  However, the upper sideband at 177 mHz displayed clear energy dependence, being detected exclusively in the 15--20 keV energy range, whereas the lower sideband at 83 mHz was observed across all selected energy bands with \astro. The non-detection of the 177 mHz feature above 20 keV is likely attributed to the increased background noise in LAXPC at higher energies.
The \nus\ observations further confirmed the energy-dependent nature of these sidebands, with the rms of the 177 mHz sideband increasing from $\sim$4\% in the 3–15 keV range to $\sim$8.6\% in the 15--30 keV range during obs 1. In obs 3, the 177 mHz sideband was detected only in the 15--30 keV range with an rms of $\sim$12\%.
These sidebands also showed time dependence, which could be due to the statistical limitations of short observation span. During the 1SD phase, \citet{Kommers98} reported both lower and upper sidebands. These sidebands were symmetric in frequency, but their power amplitude was energy-dependent. These sidebands arise due to the Fourier frequency-shifting theorem, where the QPO signal modulates the amplitude of the coherent pulsations, producing symmetric sidebands in the shape of the QPO peak around the harmonic frequencies $n\nu_{\rm NS} \pm \nu_{\rm QPO}$, where $n$=1, 2,... is an integer \citep{Kommers98}.
If the sidebands in \src\ are due to amplitude modulation of the 46 mHz QPO, there must be an additional mechanism that can explain the enhancement or reduction of the power amplitude of these sidebands in different energy ranges. 
Interestingly, the lower sideband is almost absent in weakly-magnetized neutron star binaries 4U 1608-52 and 4U 1728-34 \citep{Jonker00}. In these cases, the magnetic field influence on the accretion process is weaker, suggesting that the interaction between the neutron star's magnetic field and the accretion disc in \src\ plays a crucial role in generating the sidebands and preferentially amplifying one sideband.

\section*{Acknowledgements}
This work has made use of data from the \astro\ mission of the Indian Space Research Organisation (ISRO), archived at the Indian Space Science Data Centre (ISSDC). We thank the LAXPC Payload Operation Center (POC) 
for verifying and releasing the data via the ISSDC data archive and providing the necessary software tools. We have also made use of the software provided by the High Energy Astrophysics Science Archive Research Center (HEASARC), which is a service of the Astrophysics Science Division at NASA/GSFC. This research has also made use of \nus\ data obtained from the HEASARC and the \nus\ Data Analysis Software (NUSTARDAS) jointly developed by the ASI Science Data Center (ASDC, Italy) and the California Institute of Technology (USA). CJ acknowledges the financial assistance received from the ANRF (erstwhile SERB)–DST grant (CRG/2023/000043). AB acknowledges SERB (SB/SRS/2022-23/124/PS) for financial support and is also grateful to the Royal Society, United Kingdom. 
We also thank the anonymous referee for insightful comments and suggestions.

\section*{Data Availability}

Data used in this work can be accessed through the Indian Space Science Data Center (ISSDC) at \url{https://astrobrowse.issdc.gov.in/astro\_archive/archive/Home.jsp} and the HEASARC archive at \url{https://heasarc.gsfc.nasa.gov/cgi-bin/W3Browse/w3browse.pl}. 



\bibliographystyle{mnras}
\bibliography{sample} 




\appendix

\section{\nus\ Observations and Data Reduction}
\label{appendix:nustar}

\src\ was observed five times by \nus\ \citep{Harrison13} following the torque reversal in 2023 (Table \ref{tab:nus}). The data were processed following the procedures outlined by \citet{Sharma23c}, utilizing \textsc{heasoft} version 6.34 along with the latest calibration files (version 20250122). The calibrated and screened event files were generated using the task \textsc{nupipeline}. A circular region of radius 80 arcsec centred at the source position was used to extract the source events. Background events were extracted from a circular region of the same size away from the source. The task \textsc{nuproduct} was used to generate the science products such as light curves. These light curves were corrected to the Solar system barycentre using \textsc{barycorr}. Finally, the background-corrected light curves from the FPMA and FPMB detectors were combined using \textsc{lcmath} to obtain the summed light curves for subsequent timing analysis.

\begin{table}
    \centering
    \caption{Details of \nus\ observations, including the measured pulse period, primary QPO frequency, and corresponding rms obtained from the 3–30 keV light curve analysis.}
    \resizebox{\columnwidth}{!}{
    \begin{tabular}{cccccc}
    \hline
      Obs & Obs-ID    &  Date  & Exposure & Period & $\nu_{\rm QPO}$/rms\\
      & & (yyy-mm-dd) & (ks) & (s) & (mHz)/(\%) \\
    \hline  
      1 & 90901318002 & 2023-05-02 & 27 & 7.668021(10) & 46.8(9)/ 17(2) \\ 
      2 & 90901318004	& 2023-05-19 & 19 & 7.668057(8) & 46.9(9)/ 22(2) \\
      3 & 90901318006 & 2023-06-04 & 18 & 7.668101(5) & 46.7(8)/ 21(2) \\
      4 & 90901318008 & 2023-06-22 & 22 & 7.668135(14) & 48.2(1.1)/ 21(3) \\
      5 & 90901318010 & 2023-07-05 & 18 & 7.668171(16) & 46.3(1.1)/ 20(3) \\
    \hline     
    \end{tabular}}    
    \label{tab:nus}
\end{table}



\label{lastpage}
\end{document}